# High Dynamic Range Scanning Tunneling Microscopy

*Ajla Karić•, Carolina A Marques•, Fabian Donat Natterer\**
*•Equal contribution*

*Department of Physics, University of Zurich, Winterthurerstrasse 190, CH-8057 Zurich, Switzerland*
*\*fabian.natterer@physik.uzh.ch*

**Abstract**

We increase the dynamical range of a scanning tunneling microscope (STM) by actively subtracting dominant current-harmonics generated by nonlinearities in the current-voltage characteristics that could saturate the current preamplifier at low junction impedances or high gains. The strict phase relationship between a cosinusoidal excitation voltage and the current-harmonics allows excellent cancellation using the displacement-current of a driven compensating capacitor placed at the input of the preamplifier. Removal of DC currents has no effect on, and removal of the first harmonic only leads to a rigid shift in conductivity that can be numerically reversed by adding the known removal current. Our method requires no permanent change of the hardware but only two phase synchronized voltage sources and a multi-frequency lock-in amplifier to enable high dynamic range spectroscopy and imaging.

- Active power filter
- Dynamic range compression
- High gain preamplifier

**Graphical abstract**

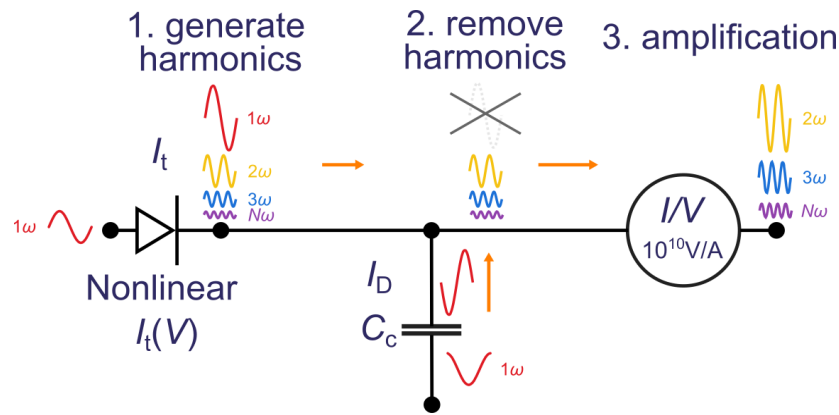

## Background

The hallmark of the scanning tunneling microscope (STM) is the exponential current-distance relationship demonstrated by Binnig and coworkers in 1982 [1], just before wowing the world with their atomic resolution imaging of the 7×7 reconstruction of the silicon surface [2]. These early milestones established the need for current preamplifiers that would span a large dynamical range. Even at constant tip-sample separation, tunneling spectroscopy quickly runs into preamplifier saturation problems when trying to simultaneously measure electronic states close to and far away from the Fermi level. This is unsatisfactory when one is interested in the local density of states, whose measurement throws out the large current background in the current derivative d$I_t(V)$/d$V$. Large bias range spectroscopy then frequently requires a lower preamplifier gain that consequently limits the minimally measurable current and thus obscures information around the Fermi level. Other STM pioneers cleverly circumvented the dynamical range limitation by synchronizing the lowering of the bias voltage with the closing of the tip-sample separation [3], which however, required an a-*posteriori* correction using the separately measured global decay length.

Here we introduce a high-dynamical range STM (HDR-STM) concept that actively compresses the dynamical range of the current, allowing us to use higher gains at small junction-impedances without saturating the preamplifier. The large gain enables us to resolve the small current amplitudes that would be lost in low gain measurements. Our approach requires no assumptions of the tunneling junction or spectroscopically synchronized tip-motion and it relies solely on the strict phase relationship between a cosinusoidal excitation voltage and the current-harmonics naturally generated by the nonlinear current-voltage characteristics $I_t(V)$. Our method is conceptually similar to capacitive displacement current compensation, but we inject displacement currents to principally remove resistive as well as capacitive currents. It can also be interpreted as a complementary active power filter that removes low-order instead of high-order harmonics.

**Method details**

The application of a cosinusoidal voltage $V(t)=V_{drv}\cos(\omega_0 t)$ on the nonlinear current-voltage characteristics $I_t(V)$ of a tunneling junction generates a time-varying tunneling current whose frequency content is composed of integer order harmonics $n\omega_0$ of the excitation frequency $\omega_0$ [4,5] (see concept in Figure 1). We write $I_t(V)$ as an infinite sum of increasing order polynomials, and apply $V(t)$ to write $I_t(V(t)) = a_0 + \sum_{n=1}^{\infty} a_n \cos^n(\omega_0 t) = b_0 + \sum_{n=1}^{\infty} b_n \cos(n\omega_0 t)$, where we used trigonometric identities[1] to see how the $n^{th}$ power transforms the excitation voltage into an $n^{th}$-order current-harmonic. This harmonic decomposition can be simulated using a pre-measured $I_t(V)$ curve, shown in Figure 1(b) for an $I_t(V)$ of Au(111). While the $b_0$ coefficient contains contributions from all even order polynomials, the $b_{n>1}$ coefficients carry contributions from polynomials smaller or equal to $n$, and they exhibit the characteristic $1/n\omega_0$ trend [see gray dashed line in Figure 1(b) right panel]. The transformation of the tunneling current into harmonics provides the basis of how their measurement can be used to reconstruct the original $I_t(V)$ characteristics [4,5]. Note that the maximally possible tunneling current of this harmonic decomposition is $I_t^{max} = \sum_{n=0}^{\infty} |b_n|$, which is dominated by the lowest integer components, $b_0$ and $b_1$. While the strong attenuation of $b_n$ of the higher order harmonics allows the faithful reconstruction of the original $I_t(V)$ curve by only measuring a finite number of the current harmonics, the large amplitudes of the DC component $b_0$ and low integer harmonics may saturate the preamplifier, requiring the use of lower gains. Unfortunately, the reduced preamplifier gain then raises the noise floor and increases the minimally detectable current. The larger noise floor also means that higher order harmonics would become obscured by noise, which would lead to a loss of information.

Fortunately, the transformation of the tunneling current into harmonics allows for the straightforward implementation of an active power filter that removes current harmonics by exploiting the perfect synchronization between the dominant (or any desired) harmonic and the cosinusoidal excitation voltage. This works by applying a synchronized voltage $V(t)=V_{cmp}\cos(n\omega_0 t+\varphi)$, with $\varphi \approx \pi/2$ and $n$ designating the targeted harmonic, onto a compensating capacitor $C_c$ placed at the input of the preamplifier. This capacitor then acts as a current source that pushes a displacement current $I_D=C_c dV/dt=-b_n$ onto the current line, perfectly opposing the respective current *before* reaching the preamplifier [Figure 1(a)]. Principally, this allows compressing the maximal current down to arbitrarily small values $I_t^{HDR} = I_t^{max} - \sum_{n=1}^{N} |b_n|$ by removing any selection of the current harmonics, [Figure 1(b)], simply by applying a superposition of properly phase shifted and scaled integer harmonics of the excitation voltage on $C_c$. This harmonic removal then enables operating the preamplifier at the maximal possible gain to substantially increase the dynamical range and signal-to-noise ratio of the STM. The example of Figure 1(c) shows the reduction of the maximal current ($I_{max}$) amplitudes during spectroscopy when the DC component ($b_0$) or $b_1$ are removed [see also zoom in Figure 1(d)]. Since $I_{max}$ determines the saturation of the preamplifier, it becomes immediately clear that the compressed spectrum could now be measured at a higher gain.

At first sight, the removal of current harmonics might appear unacceptable because it seemingly discards information about the tunneling junction, but we recall that we have perfect knowledge of these missing currents as they are equivalent to their compensating displacement currents. Information is irreversibly lost, however, if one removed the zero-frequency component $b_0$, for instance by using the (optional) DC blocking capacitor $C_b$ at the preamplifier input. The latter might be acceptable, if the researcher was mostly interested in the conductivity rather than absolute current

---

[1] For even $n$: $\cos^n\Theta = \frac{1}{2^n}\binom{n}{\frac{n}{2}} + \frac{2}{2^n}\sum_{k=0}^{\frac{n}{2}-1}\binom{n}{k}\cos((n-2k)\Theta)$ and for odd $n$: $\cos^n\Theta = \frac{2}{2^n}\sum_{k=0}^{\frac{n-1}{2}}\binom{n}{k}\cos((n-2k)\Theta)$

values, for instance when focusing on d$I_t$(V)/dV, such as for quasiparticle interference measurements [6]. The effect of $b_0$ and $b_1$ removal on the local density of states measured via d$I_t$(V)/dV can be seen in Figure 1(e). Removing $b_1$ only leads to a rigid shift towards negative LDOS while preserving the overall shape of the spectrum.

**Method validation**

We are now looking at the experimental implementation of our active power filter concept for HDR-STM. Our method requires only the reversible addition of the compensating capacitor $C_c$ (1 pF) placed via a tee-piece at the input of the preamplifier and the optional series DC blocking capacitor $C_b$ (100 pF) that removes the $b_0$ component of the current [Figure 1(a)]. To generate and compensate the current harmonics -$b_n$, we use a synchronized dual channel waveform generator, implemented in a multifrequency lock-in amplifier (Intermodulation products, MLA-3) whose first output acts as the excitation voltage $V(t)=V_{drv}\cos(\omega_0 t)$ and the second output $V(t)=V_{cmp}\cos(\omega_0 t+\varphi)$ in combination with $C_c$ as the harmonic-cancelling current source. The tunneling current is fed into a large bandwidth preamplifier (NF corp SA-606F2: gain $10^9$ V/A or NF corp SA-607F2: gain $10^{10}$ V/A), whose signals provide current information for STM operation and for harmonic demodulation at the lock-in amplifier. The preamplifier bandwidth determines the frequency of the highest harmonic that may still reach the lock-in amplifier, which in our case is 40'320 Hz for the 63$^{rd}$ harmonic of our preferred excitation frequency of $f$=640 Hz.

We have previously used the compensating capacitor $C_c$ to accurately cancel out the stray displacement current $I_D$ using $V(t)=V_{drv}C_c/C_s\cos(\omega_0 t+\varphi_c)$, with $\varphi_c\approx\pi$ [5] to produce an opposing current. In addition to that stray capacitive current compensation, we now also apply $V_{cmp}=V_{cmp}\cos(\omega_0 t+\varphi)$, $\varphi=\varphi_c-\pi/2$ onto $C_c$ to deliberately generate -$b_n=C_c dV(t)/dt$ that eliminates the strong harmonics generated by the $I_t$(V) curve. Since capacitive and resistive currents have a π/2 phase shift, we first compensate the capacitive displacements to fix the phase relationship. Our setup requires no modification other than the application of suitably large cancellation voltages or adjustments to the compensating capacitance $C_c$. Our presently used $C_c$=1 pF and the 12 V maximal excitation amplitude, allows current compensation up to about 50 nA. Accordingly, larger currents of small junction impedances could be compensated by increasing $C_c$.

We next validate our HDR-STM method for tunneling spectroscopy, notably to verify the benign action of the harmonic removal on local density of states (LDOS) measurements. Using $V_{drv}$=600 mV, we start by examining the height-dependence of the harmonic removal, characterized by Δ$b_1$ as shown in Figure 2(a). The blue line shows Δ$b_1$ without active compensation (-$b_1$=0) as it follows the same exponential distance-dependence that is found for the tunneling current. This exponential trend allows $b_1$ to be used for feedback control of the tip-height [7], which may be useful to safely scan the surface even without DC current or when the time-averaged current would be zero for nearly ohmic or electron-hole symmetric junctions. When we apply a removal current -$b_1$ via $C_c$, the height dependence of Δ$b_1$ (yellow) shows the action of harmonic removal by minimizing Δ$b_1$ at an ideal tip height where the resistive part $b_1$ exactly matches the removal current, showing the excellent cancellation potential of our HDR-STM method. In addition, in the height range close to this ideal tip-height, Δ$b_1$ is extremely sensitive to height-deviations, indicating the potential for phase sensitive feedback control (not shown). Figure 2(b) shows the reconstructed $I_t$(V) curves measured on Au(111) with varying magnitude of $b_1$ removed using a compensation current of 31.2 nA, which would saturate the $10^9$ gain of the preamplifier. The current with active harmonic removal shows the same shape as found in the simulation of Figure 1(d). The closer the resistive and removal currents match (Δ$b_1$→0), the smaller is $I_{max}$, and the more compressed is the spectrum. The crossing point of the three curves corresponds to the zero-bias condition and it can accordingly be used to quantify bias-offsets present in the system. If we additionally removed $b_0$, the compressed $I_t$(V) curve would center around the zero current mark, further reducing $I_{max}$, as also seen in Figure 1(d). The LDOS in Figure 2(c) clearly shows the onset of the characteristic Au(111) surface state at about -500 mV as it does for conventionally measured spectra using much smaller effective currents. The removal of $b_1$ only rigidly shifts the LDOS but preserves the spectral details, as can be seen from the coinciding wiggles among the three d$I_t$(V)/dV traces.

Figure 3(a) shows a topographic image of Au(111), measured with a $C_b$=100 pF DC blocking capacitor ($b_0$=0), $V_{drv}$=600 mV and a feedback loop maintaining constant log(Δ$b_1$). In this mode, there is no more galvanic connection between the tip and the preamplifier circuit. Figure 3(b) shows a topographic scan while actively removing a large part of the resistive current carried by the first harmonic. This scan operates by setting the feedback controller to maintain

log($\Delta b_1$)=300 pA while having -$b_1$=36 nA active. We have also achieved atomic resolution in HDR-STM mode, as shown in Figure 1(a), which we measured at -$b_1$≈50 nA and at a gain of $10^{10}$ or equivalent to a dynamical range of 146 dB[2]. Our specified current estimate is here approximate because the gain limited the maximal current to 1 nA, which was occasionally exceeded with the second order harmonic.

**Limitations**

When implementing the high dynamic range STM method, we encountered a few limitations that we discuss in the following.

The first is the large value of the DC component $b_0$ that can grow to the same order of magnitude than the first harmonic [see simulation in Figure 1(b)], because all even order polynomials contribute to $b_0$ via the polynomial decomposition mentioned above. The large value of $b_0$ has motivated our use of the blocking capacitor $C_b$ in parts of the experimental verification, despite its irreversibility on the current reconstruction. For future studies related to quasiparticle interference, we may be able to accept this loss of information because those studies will mostly focus on the spatial dependence of d$I_t$($V$)/d$V$. One remedy for a large $b_0$ could be to add a DC voltage $V(t)=V_{DC}+V_{drv}\cos(\omega_0 t)$ that maximizes the amplitude of odd order harmonics, which can be simulated using conventionally measured spectra.

A second limitation is the cancellation of the current only after transfer across the tunneling junction. This may lead to noticeable Joule heating in extreme cases. For instance, the impedances used by Binnig [1] and Limot [8], would dissipate a power equivalent of order 1 µW. The Feenstra spectroscopy method [9] would then be a suitable alternative but its normalization protocol requires the measurement of the decay length for every spectroscopy location and the consideration of potential tip-gating effects [10].

A third limitation follows from the use of high-gain preamplifiers that have a reduced bandwidth and that either curtail the number of harmonics that can be demodulated or requires the use of lower excitation frequencies with accordingly longer spectroscopy times. High gain preamplifiers have not only a smaller bandwidth, they also require the application of several compensating voltages because the amplitudes of higher order current harmonics may still be too strong and saturate the preamplifier, as was partially the case in the topographic scan of Figure 1(a). When more harmonics are removed, fewer demodulators are available for the measurement of information carrying resistive harmonics.

The noise that couples into the system via the compensation capacitor is also a limitation that needs attention since any signal transferred onto the current line will be amplified by the gain of the preamplifier. We observed the presence of noticeable switching noise related noise peaks around 8 kHz when we changed the compensation capacitor from 1 pF to 10 pF. Likewise, harmonic distortion in the cosinusoidal compensation voltage might produce unwanted compensation currents. A potential remedy would be suitable low-pass filters, bandpass filters that only transmit the compensation frequency, and radiofrequency filters.

**CRediT author statement**

*Ajla Karic: Validation, Investigation, Funding acquisition. **Carolina A. Marques**: Investigation, Supervision, Funding acquisition. **Fabian D. Natterer**: Conceptualization, Methodology, Validation, Investigation, Writing – original draft, Supervision, Funding acquisition.*

**Acknowledgments**

Funding: This work was supported by the Swiss National Science Foundation [200021_200639, PP00P2_211014, 206021_213238, CRSK-2_221052] and through the Federal Commission for Scholarships for Foreign Students for the Swiss Government Excellence Scholarship (ESKAS No. 2023.0017 and ESKAS No. 2024.0221). We thank Achim Vollhardt for fruitful discussions.

**Declaration of interests**

☒ The authors declare that they have no known competing financial interests or personal relationships that could have appeared to influence the work reported in this paper.

☐ The authors declare the following financial interests/personal relationships which may be considered as potential competing interests:

---

[2] Dynamic range =20log(50 nA/2.5 fA)=146 dB, where 2.5 fA represents the preamplifier noise level

## References

[1] G. Binnig, *Tunneling through a Controllable Vacuum Gap*, Appl. Phys. Lett. **40**, 178 (1982).

[2] G. Binnig, H. Rohrer, Ch. Gerber, and E. Weibel, *7 x 7 Reconstruction on Si(111) Resolved in Real Space*, Phys. Rev. Lett. **50**, 120 (1983).

[3] P. Mårtensson and R. M. Feenstra, *Geometric and Electronic Structure of Antimony on the GaAs(110) Surface Studied by Scanning Tunneling Microscopy*, Phys. Rev. B **39**, 7744 (1989).

[4] R. Borgani, M. Gilzad Kohan, A. Vomiero, and D. B. Haviland, *Fast Multifrequency Measurement of Nonlinear Conductance*, Phys. Rev. Applied **11**, 044062 (2019).

[5] B. Zengin, J. Oppliger, D. Liu, L. Niggli, T. Kurosawa, and F. D. Natterer, *Fast Spectroscopic Mapping of Two-Dimensional Quantum Materials*, Phys. Rev. Research **3**, L042025 (2021).

[6] L. Petersen, Ph. Hofmann, E. W. Plummer, and F. Besenbacher, *Fourier Transform–STM: Determining the Surface Fermi Contour*, Journal of Electron Spectroscopy and Related Phenomena **109**, 97 (2000).

[7] H. Alemansour, S. O. R. Moheimani, J. H. G. Owen, J. N. Randall, and E. Fuchs, *Ultrafast Method for Scanning Tunneling Spectroscopy*, Journal of Vacuum Science & Technology B **39**, 042802 (2021).

[8] L. Limot, T. Maroutian, P. Johansson, and R. Berndt, *Surface-State Stark Shift in a Scanning Tunneling Microscope*, Phys. Rev. Lett. **91**, 196801 (2003).

[9] R. M. Feenstra, *Tunneling Spectroscopy of the (110) Surface of Direct-Gap III-V Semiconductors*, Phys. Rev. B **50**, 4561 (1994).

[10] Y. Zhao, J. Wyrick, F. D. Natterer, J. F. Rodriguez-Nieva, C. Lewandowski, K. Watanabe, T. Taniguchi, L. S. Levitov, N. B. Zhitenev, and J. A. Stroscio, *Creating and Probing Electron Whispering-Gallery Modes in Graphene*, Science **348**, 672 (2015).

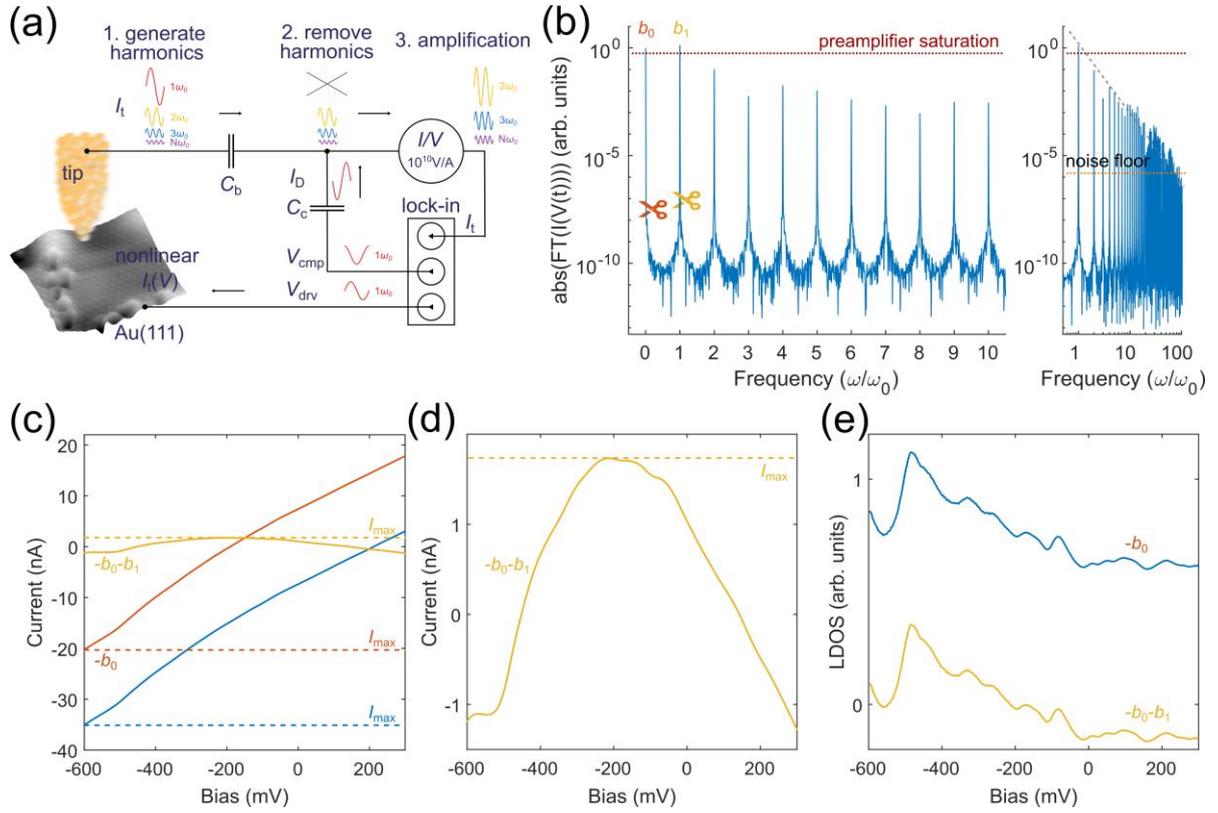

*Figure 1: High dynamic range scanning tunneling microscopy concept*. **(a)** The application of a harmonic excitation $V_{drv}=V_{drv} \cos(\omega_0 t)$ on the nonlinearities $I_t(V)$ of the tunneling junction creates higher order current harmonics that can be subtracted by a tailored displacement current $I_D$ created with the compensating capacitor $C_c$ and the application of $V_{cmp}$. The optional DC blocking capacitor $C_b$ removes the DC current component ($b_0$) to further increase the dynamical range of the STM. The atomically resolved Au(111) surface was measured without DC ($b_0$) and first harmonic current ($b_1$) at preamplifier gain of $10^{10}$ at an effective current of about 50 nA ($V_{drv}$=150 mV, f=640 Hz, $-b_1 \approx$50 nA, T=4.3 K). **(b)** Simulated harmonic decomposition of $I_t(V)$ of Au(111) after excitation with $V(t)=V_{drv} \cos(\omega_0 t)$ using a conventionally measured spectrum. The removal of high amplitude current harmonics (scissor symbols) prevents preamplifier saturation (dotted red line). The panel on the right shows the harmonic decomposition up to higher order harmonics, where the dashed gray line shows the $1/\omega$ trend of the higher order harmonics and the dotted orange line indicates the noise level of the current preamplifier. A smaller tip-sample distance would lift more harmonics above the noise level. **(c)** $I_t(V)$ curves created using all harmonics (blue), without DC component ($b_0$, red), and without DC ($b_0$) and first harmonic ($b_1$, yellow). The latter shows the strongest compression of the current range, enabling the use of larger preamplifier gains. **(d)** Zoom into compressed current from (c). **(e)** The removal of the DC component ($b_0$) has no effect on the local density of states (LDOS $\propto dI_t(V)/dV$, blue), and the removal of the first harmonic only rigidly shifts the LDOS (yellow). The original $I_t(V)$ and LDOS could be reconstructed using the known amplitude of $b_1$ while the removal of the DC component ($b_0$) only impacts $I_t(V)$, albeit irreversibly.

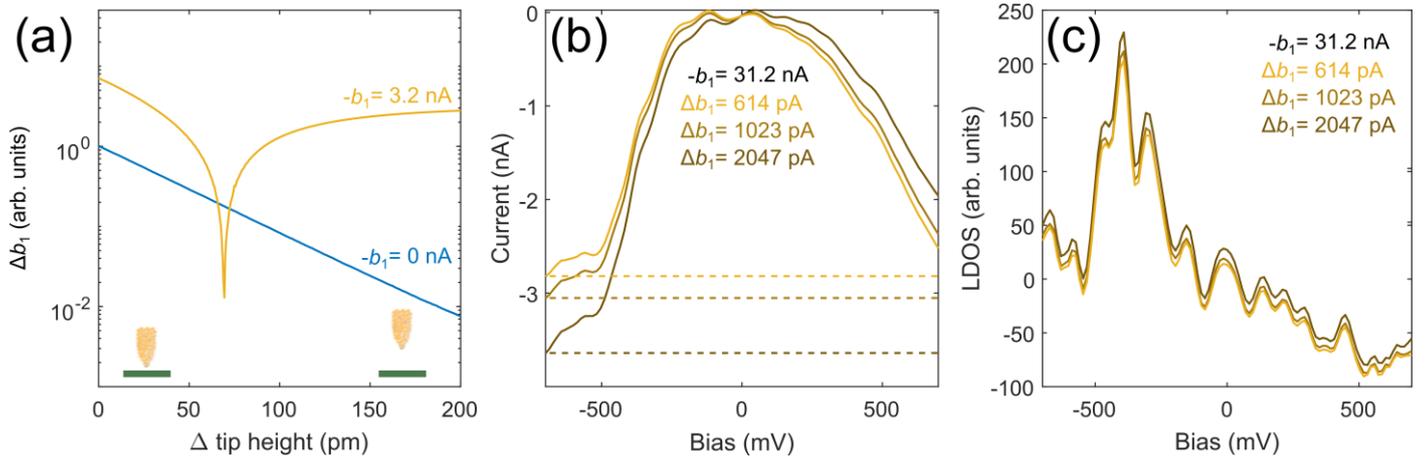

*Figure 2: **High dynamic range spectroscopy.** (a)* Tip height dependence of the difference ($\Delta b_1$) between the amplitude of the first harmonic and the current from active compensation ($b_0=0$, $V_{drv}=600$ mV, $f=640$ Hz, gain $10^9$ V/A, $T=4.3$ K). Without harmonic removal ($-b_1=0$), the amplitude of the first harmonic follows an exponential trend (blue). With active harmonic compensation ($-b_1=3.2$ nA), $\Delta b_1$ (yellow) shows a minimum that corresponds to the tip-height at which the best compensation is achieved and the active harmonic removal almost fully suppresses $b_1$. *(b)* $I_t(V)$ spectroscopy of Au(111) with active harmonic removal of $-b_1=31.2$ nA at different tip-heights. At small $\Delta b_1$, the current-range is more compressed and the resistive part better matches the active compensation current $-b_1$ ($-b_1=31.2$ nA, $V_{drv}=600$ mV, $f=640$ Hz, gain $10^9$ V/A, $T=4.3$ K). The DC component was not removed ($b_0\neq 0$) and the dashed lines mark the respective maximal currents $I_{max}$. *(c)* $dI_t(V)/dV$ spectra with active compensation, numerically differentiated from (b). The LDOS shifts rigidly towards negative values, the better we remove $b_1$ using active compensation. Importantly, the harmonic removal preserves the spectral structures, as can be seen by the coincidence of the LDOS features between the three different traces.

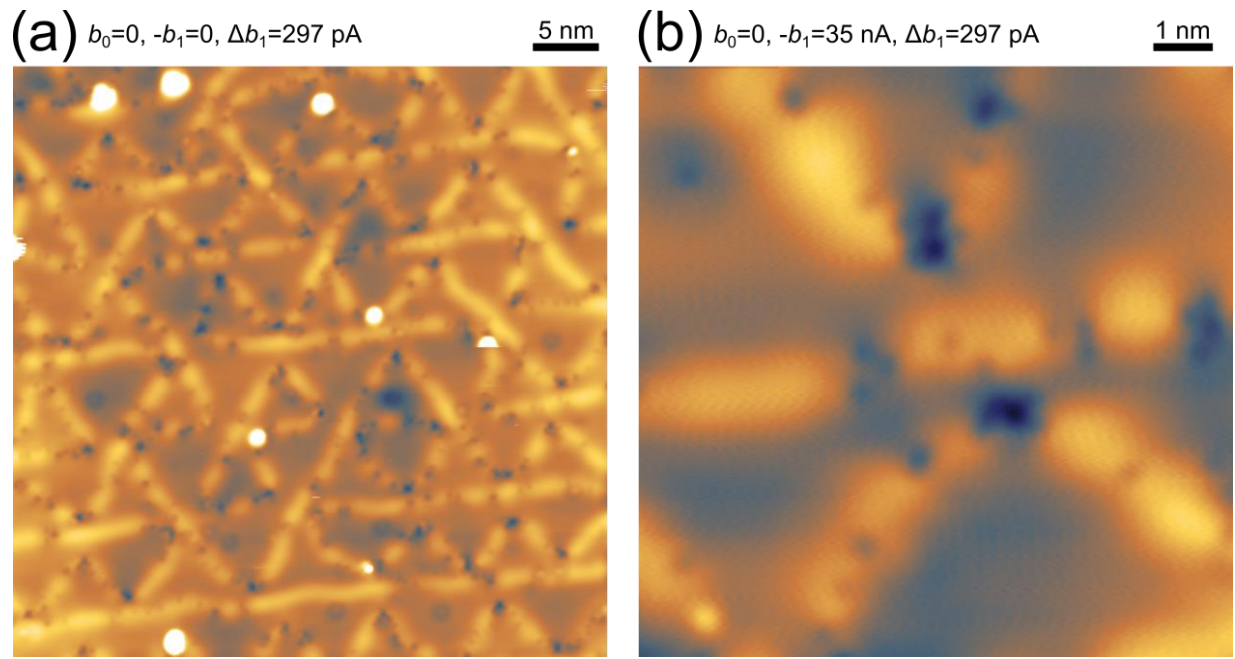

*Figure 3: **Topographic measurement using HDR-STM**. **(a)** Topography of Ag(111) without DC current ($b_0$=0) and feedback loop on log($\Delta b_1$) ($V_{drv}$=600 mV, $\Delta b_1$=297 pA, f=640 Hz, gain $10^9$ V/A, T=4.3 K). **(b)** Closed loop topography to maintain $\Delta b_1$=297 pA, while actively subtracting -$b_1$=36.45 nA ($b_0$=0, $V_{drv}$=600 mV, f=640 Hz, gain $10^9$ V/A, T=4.3 K).*